# Weather Forecasting using Incremental K-means Clustering

SANJAY CHAKRABORTY
National Institute of Technology
(NIT) Raipur, CG, India.
email: sanjay_ciem@yahoo.com

Prof. N.K.NAGWANI
National Institute of Technology
(NIT) Raipur, CG, India.
email:nknagwani@gmail.com

LOPAMUDRA DEY
University of Kalyani
Kalyani, W.B., India
email: lopamudra.dey1@gmail.com

*Abstract* – Clustering is a powerful tool which has been used in several forecasting works, such as time series forecasting, real time storm detection, flood forecasting and so on. In this paper, a generic methodology for weather forecasting is proposed by the help of incremental K-means clustering algorithm. Weather forecasting plays an important role in day to day applications. Weather forecasting of this paper is done based on the incremental air pollution database of west Bengal in the years of 2009 and 2010. This paper generally uses typical K-means clustering on the main air pollution database and a list of weather category will be developed based on the maximum mean values of the clusters. Now when the new data are coming, the incremental K-means is used to group those data into those clusters whose weather category has been already defined. Thus it builds up a strategy to predict the weather of the upcoming data of the upcoming days. This forecasting database is totally based on the weather of west Bengal and this forecasting methodology is developed to mitigating the impacts of air pollutions and launch focused modeling computations for prediction and forecasts of weather events. Here accuracy of this approach is also measured.

## I. INTRODUCTION

Forecasting is very important for prediction of the future events. Science and computer technology together has made significant advances over the past several years and using those advanced technologies and few past patterns, it grows the ability to predict the future. Weather forecasting is directly dependent with the characteristics of the particulate matters present in the air.

Weather forecasting ← max (effects of $NO_2$ + $SO_2$ + $CO_2$ + RPM +……) (1)

This paper presents a methodology for forecasting weather of 'West Bengal' through clustering. This methodology uses an air pollution database which is described in later section. The paper is organized into the six sections. Section two is consisting of work done previously in the same directions; a brief background history is covered in this section. The methodology is explained elaborately in section three. The proposed model is explained in section four; here various stages are explained to perform weather forecasting using incremental K-means clustering. Section five is simulation result; here the proposed technique is applied over the test dataset and results are captured. And the last section i.e. section six is conclusion and future scope of the proposed work.

## II. RELATED WORK

There are several approaches that have been used for weather prediction. In some cases, advance numerical analysis has used for weather prediction but in most of the situations clustering techniques are used for different types of predictions. It may be weather prediction or may be natural disaster prediction. All of these researches help to survive the world from the natural destructive events. Weather forecasting can also be done by using artificial neural network [9].

This paper is based on the incremental approach of K-means clustering algorithm which has been already developed and discussed [1] [2]. Based on that incremental algorithm concepts weather prediction of 'West Bengal' is done in this paper. There are also exist several approaches which provide some modifications of this algorithm [3][4][5][6]. An approach is proposed on a case study of time series forecasting through clustering. In this approach, a generic methodology for time series forecasting is proposed. This methodology first search some useful patterns in the form of curves and it then facilitates the forecasting through linear regression by matching to the closest pattern to each time series that has to be predicted. This approach is applied on Kddcup 2003 dataset [7]. Some work is done on real time storm detection through data mining. In this approach, a model and algorithms for bridging the gap between the physical environment and the cyber infrastructure framework by means of an events processing approach to responding to anomalous behavior and sophisticated data mining algorithms that apply classification techniques to the detection of severe storm patterns. The above ideas have been implemented in the LEAD-CI prototype [8]. There exists one approach which presents the data mining activity that was employed to mining weather data. The self-organizing data mining approach employed is the enhanced Group Method of Data Handling (e-GMDH). The weather data used for the DM research include daily temperature, daily pressure and monthly rainfall. Experimental results indicate that the above approach is useful for data mining technique for forecasting weather data [10].

## III. METHODOLOGY

This analysis is based on the observation of the air pollution data has been collected from the "West Bengal Air Pollution Control Board" and the URL is- "http://www.wbpcb.gov.in/html/airqualitynxt.php". This database consists of four air-pollution elements or attributes and they are Carbon dioxide ($CO_2$), Respirable particulate matter (RPM), Sulphur dioxide ($SO_2$) and Oxides of Nitrogen ($NO_x$). Air pollution data of each day are collected and stored that record in an .arff (Attribute resource file format) file format. The detail database format is shown in the 'Table 1'.

Table I. Original air-pollution Database

| *Date* | *$CO_2$* | *RPM* | *$SO_2$* | *$NO_X$* |
|---|---|---|---|---|
| 1/1/2009 | 85 | 183 | 12 | 95 |
| 2/1/2009 | 95 | 289 | 14 | 125 |
| 3/1/2009 | 112 | 221 | 10 | 101 |
| 4/1/2009 | 114 | 191 | 11 | 97 |
| 5/1/2009 | 100 | 175 | 11 | 101 |
| 6/1/2009 | 78 | 149 | 7 | 93 |
| ………. | ……. | ……. | …… | …….. |
| 1/2/2009 | 120 | 197 | 10 | 105 |
| 2/2/2009 | 115 | 151 | 10 | 85 |
| 3/2/2009 | 98 | 154 | 8 | 96 |
| 4/2/2009 | 90 | 195 | 8 | 93 |
| ………. | ……. | ……. | …… | …….. |

The above database is a dynamic database where data are updated frequently. The main approach of this paper is that first apply the K-means clustering algorithm [Chakraborty and Nagwani, 2011] on that above original database (assuming initial cluster number). Then compute the means of each cluster based on their air polluted attributes in the database. Then a list of weather category will be developed based on the maximum mean value of each cluster. Now when the new data (data of upcoming days) are inserted into the old database and then apply the incremental K-means clustering algorithm. Based on the behaviour of the incremental K-means clustering algorithm the minimum means of the new cluster data can be computed and it can be easily defined to which cluster the new means are belonged and their weather category can also be defined based on that particular cluster's weather category. At last the accuracy of this method is also measured and discussed.

*A. Effects of air-pollution data on weather*

The experimented database of this paper consists of four air-polluted data ($CO_2$, RPM, $SO_2$, $NO_X$). This four data has very important roles in the weather or climate change. They not only make impact on the climate but they can also harmful for humans and plants. These air pollutants are directly emitted from several different sources, such as ash from a volcanic eruption, gas from a motor vehicle exhaust, released from various industrial processes, high temperature combustion and so on.

The following 'Fig.1' shows how the air pollution data affect on humans health, plants health and moreover on the environment [11].

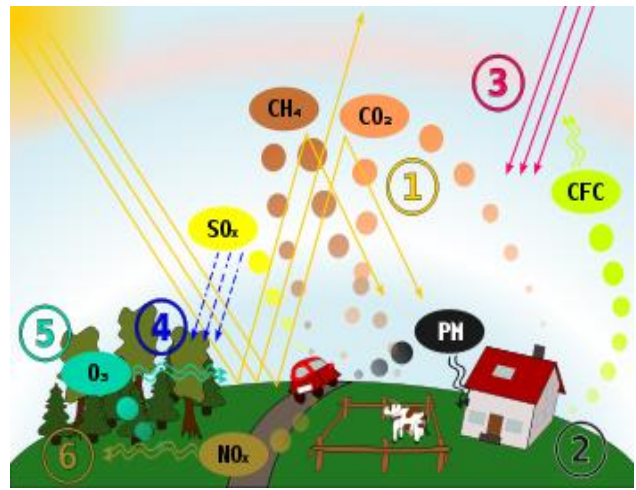

Figure.1 Effects of air pollution data on humans, plants and environment

Their effects on climate change are listed below,

- **$CO_2$:** A colourless, odorless, non-toxic greenhouse gas associated with ocean acidification, emitted from sources such as combustion, cement production, and respiration. It is one of the main pollutants of causing 'global warming'. Due to the 'greenhouse effect', the temperature of the environment is increased and most importantly seasonal change is caused due to the increase of $CO_2$ [11].
- **$SO_2$:** $SO_2$ is produced by volcanoes and in various industrial processes. Since coal and petroleum often contain sulphur compounds, their combustion generates sulphur dioxide. Further oxidation of $SO_2$, usually in the presence of a catalyst such as $NO_2$, forms $H_2SO_4$, and thus acid rain. $SO_2$ also form smog (smoke+fog) which create visibility problem [11][12][13].

- **$NO_X$:** Nitrogen dioxide is emitted from high temperature combustion. It can be seen as the

brown haze dome above or plume downwind of cities. $NO_2$ is one of the most prominent air pollutants. Just like $CO_2$, $NO_2$ is responsible to increase temperature and it also creates smog [11][12][13]..

- **RPM:** Particulates, alternatively referred to as particulate matter (PM) or fine particles, are tiny particles of solid or liquid suspended in a gas. It creates dust, smokes, fumes, mist, fog, aerosols, fly ash and so on. Increased levels of fine particles in the air are linked to health hazards such as heart disease, altered lung function and lung cancer [11][12][13].

*B. Mathematical Explanation*

Suppose there is a set of air pollution data which consist of 15 data. Suppose each data represents the data of each day. The set is shown by the table below,

Table II. Sample air pollution database

| Air Pollutant data | | | |
|---|---|---|---|
| $CO_2$ | RPM | $SO_2$ | $NO_X$ |
| 82 | 14 | 12 | 24 |
| 72 | 56 | 28 | 8 |
| 36 | 2 | 48 | 5 |
| 7 | - | 94 | 62 |

Here typical K-means is applied for initial data and incremental K-means [1] is applied for incremental or new coming data. Let assume, initially the value of clusters is 4 (K=4) & initially the means of those four clusters are $C_1=8, C_2=56, C_3=28, C_4=72$ and also assume the above database contains no noisy data.

*First Iteration:*
Now first apply typical K-means on the above data by using Manhattan distance metric ($|A_i - A_j|$),
Now for the first data 12($SO_2$),
$C_1 = |8 - 12| = 4$ (minimum)
$C_2 = |56 - 12| = 44$
$C_3 = |28 - 12| = 16$
$C_4 = |72 - 12| = 60$
So, 12 ⟶ $C_1$
Thus apply the same technique for the above all data,

|  | #items | Means |
|---|---|---|
| $C_1 = \{12, 8, 5, 14, 7, 2\} =$ | 6 | 8 |
| $C_2 = \{56, 48, 62\}$ = | 3 | 55.33 |
| $C_3 = \{28, 24, 36\}$ = | 3 | 29.33 |
| $C_4 = \{72, 82, 94\}$ = | 3 | 82.66 |

*Second Iteration:*
Now, again perform clustering based on the above new generated means,
$C_1=12$   $C_2=48$   $C_1=5$   $C_2=62$
$C_3=24$   $C_1=8$    $C_1=14$  $C_1=7$
$C_4=82$   $C_4=72$   $C_3=28$  $C_1=2$
$C_3=36$   $C_4=94$   $C_2=56$

The result of the second iteration is same as the above.
From the above four clusters the nature of those clusters on climate change can be measured.

- From the resultant data of cluster1($C_1$), it can be said that the effect of **RPM** (14 is maximum) are more compare to the other pollutants. So, as per their effects on weather (discussed above), the weather of those particular days were smogy in nature and also lots of dust, fly ash was there in the weather.

- As per the nature of Cluster2($C_2$), the weather of those particular days were hot, dry and smogy in nature due to the effect of $NO_x$.

- As per the nature of Cluster3($C_3$), the weather of those particular days were hot, smogy and humid due to effect of $CO_2$(**'Greenhouse effect'**).

- As per the nature of Cluster4($C_4$), the weather of those particular days were hot, smogy and also there may be chance of acid rain due to the effect of $SO_2$.

Now some new data are inserted into the existing database means air pollution data of some upcoming days are inserted, such as 49($NO_x$), 78($SO_2$), 20($CO_2$).
Here incremental K-means clustering algorithm [1] can be applied, according to the incremental K-means algorithm the new data are directly cluster by using direct means calculation between the new data and the means of the existing clusters. There is no need to run the whole algorithm again and again. Then the expected result is,

i. $C_1 = |8 - 49| = 41$
   $C_2 = |55.33 - 49| = 6.33$ (minimum)
   $C_3 = |29.33 - 49| = 19.67$
   $C_4 = |82.66 - 49| = 33.66$
That's why, 49($NO_x$) ⟶ $C_2$
So, the data 49 follows the features of cluster2 and it indicates that the weather of the next day will be hot and smogy in nature.

ii. $C_1 = |8 - 78| = 70$
    $C_2 = |55.33 - 78| = 22.67$
    $C_3 = |29.33 - 78| = 48.67$
    $C_4 = |82.66 - 78| = 4.66$ (minimum)
That's why, 78($SO_2$) ⟶ $C_4$
So, the data 78 follows the features of cluster4 and it indicates that the weather of that day will be hot, smogy and also there may be chance of acid rain.

iii. $C_1 = |8 - 20| = 12$
     $C_2 = |55.33 - 20| = 35.33$
     $C_3 = |29.33 - 20| = 9.33$ (minimum)
     $C_4 = |82.66 - 20| = 62.66$
That's why, 78($SO_2$) ⟶ $C_3$

So, the data 20 follows the features of cluster3 and it indicates that the weather of that day will be hot, smogy and humid due to the effect of $CO_2$.

## IV. PROPOSED MODEL

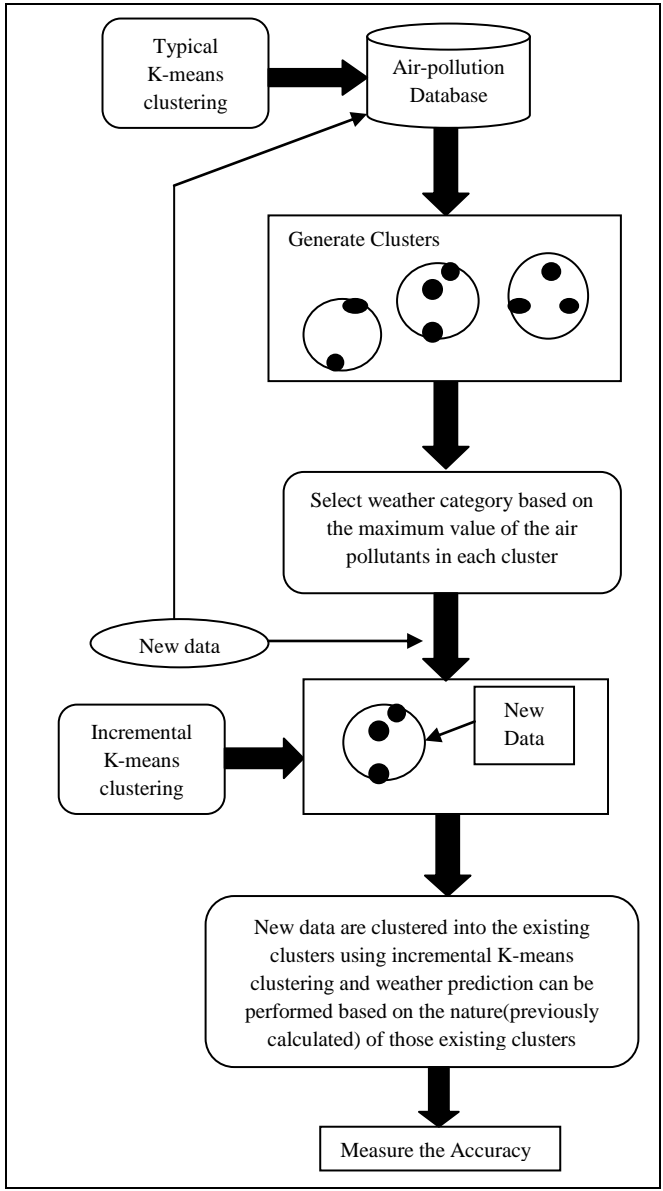

Figure.2 Proposed model of weather forecasting using incremental K-means clustering

From 'Fig.2' it can be shown that first apply typical K-means clustering on the air pollution database and based on the maximum value of the air pollutants the weather category can be defined. Now, when the new data are inserted into the existing database, then the new data are directly clustered into those existing clusters which weather category has been already decided. This new data insertion can be done using incremental K-means clustering algorithm. So, the weather category of those new data can be evaluated from the weather category of those clusters where the new data belong.

## V. SIMULATION RESULT

The simulation is totally based on the data of the year 2009 and 2010. This simulation is going to calculate the accuracy of the approach of this paper. This experiment is done with the help of Java, Weka software and it performs on the 2.26 GHz Core i3 processor computer with 4GB memory, running on Windows 7 home basic. Accuracy of any method can be measured by compare the actual value with the current value of the new method.

$$\text{Accuracy} = \frac{Number\ of\ mactched\ records}{Total\ number\ of\ records} \times 100 \quad (2)$$

The result is shown below after applying typical K-means on the air-pollution database (initially it contains the data of the first 8 months of the year 2009),

Table III. Means of initial clusters

| clusterid | clustCO$_2$mean | clustRPMmean | clustSOmean | clustNOmean |
|---|---|---|---|---|
| cluster0 | 221.376238 | 110.366337 | 10.128713 | 92.415842 |
| cluster1 | 112.600000 | 118.562500 | 8.425000 | 72.187500 |
| cluster2 | 39.458824 | 36.176471 | 6.158824 | 41.523529 |
| cluster3 | 65.196721 | 75.983607 | 7.704918 | 57.04918 |
| cluster4 | 225.943182 | 145.022727 | 12.034091 | 107.10227 |

Based on that above means, five clusters are produced and the nature of each cluster depends upon the maximum value of the mean attribute. Such as, for the first cluster the value of $CO_2$ mean is maximum, now if the new coming data of the upcoming day is inserted into the first cluster means the weather of that particular day is hot, smogy and humid due to the effect of $CO_2$.

Suppose the data(10 months) from the September month of the year 2009 to June month of the year 2010 is shown by the 'Table 5',

Table IV. Weather category according to cluster data

| Cluster Number | Cluster Maximum data | Weather Category |
|---|---|---|
| Cluster0 | 221.376238 | hot, smogy and humid |
| Cluster1 | 118.562500 | dusty, fly ash, smogy, fog, Mist |
| Cluster2 | 41.523529 | Hot, dry and smogy |
| Cluster3 | 75.983607 | dusty, fly ash, smogy, fog, Mist |
| Cluster4 | 225.943182 | hot, smogy and humid |

Table V. Pollution data of the year of 2009(September) and 2010

| Date | CO$_2$ | RPM | SO$_2$ | NO$_X$ |
|---|---|---|---|---|
| 1/9/2009 | 66 | 27 | 5 | 31 |
| 2/9/2009 | 27 | 83 | 5 | 36 |
| 3/9/2009 | 88 | 30 | 5 | 35 |
| 4/9/2009 | 98 | 29 | 5 | 35 |
| 5/9/2009 | 74 | 28 | 5 | 33 |
| ………. | ……. | ……… | …… | ……… |
| 28/9/2009 | 116 | 43 | 6 | 52 |
| 29/9/2009 | 125 | 53 | 6 | 60 |

| 30/9/2009 | 188 | 100 | 7 | 67 |
|---|---|---|---|---|
| ………. | ……. | ……. | …… | …….. |
| 1/1/2010 | 200 | 150 | 12 | 107 |
| 2/1/2010 | 220 | 160 | 13 | 110 |
| ………. | ……. | …….. | ……. | ………. |
| 1/3/2010 | 260 | 170 | 14 | 105 |
| 2/3/2010 | 270 | 175 | 14 | 112 |
| ………. | ……. | …….. | ……. | ……. |
| 1/6/2010 | 190 | 145 | 16 | 120 |
| 2/6/2010 | 200 | 155 | 12 | 118 |
| ………. | ……. | ……. | ……… | …….. |

Now the new data are inserted into the existing database and incremental K-means clustering is used to cluster those data, The new data after proper clustering by the incremental clustering algorithm it will insert into one of those existing clusters whose weather category has already been defined in the 'Table 4', then the result of forecasting is shown below from the month of September, 2009 to June 2010 (note that the database contains all the data of the West Bengal pollution board of the year 2009 and 2010). Here for calculation 'Euclidean metric' is used. Examples of the first three new calculated data is given below,

i. Cluster0= 186.7885
   Cluster1= 110.7402
   Cluster2= 29.8850 (minimum)
   Cluster3= 55.5580
   Cluster4= 212.9605

So, the incremental data of the second day (1/9/2009) should be inserted into the 'Cluster2', it follows the same nature like 'Cluster2'.

ii. Cluster0= 204.29954
   Cluster1= 99.5656
   Cluster2= 48.78
   Cluster3= 44.2561(minimum)
   Cluster4= 220.295

So, the incremental data of the second day (2/9/2009) should be inserted into the 'Cluster3', it follows the same nature like 'Cluster3'.

iii. Cluster0= 166.0447
   Cluster1= 99.2124
   Cluster2= 49.3790(minimum)
   Cluster3= 55.9282
   Cluster4= 193.666

So, the incremental data of the third day (3/9/2009) should be inserted into the 'Cluster2', it follows the same nature like 'Cluster2'. Thus the same way other data of the 10 months (from September 2009 to June 2010) are calculated.

Table VI. Weather forecasting from September, 2009 to June 2010

| Date | New data inserted into | Weather Category |
|---|---|---|
| 1/9/2009 | Cluster2 | Hot, dry and smogy |
| 2/9/2009 | Cluster3 | dusty, fly ash, smogy, fog, Mist |
| 3/9/2009 | Cluster2 | Hot, dry and smogy |
| 4/9/2009 | Cluster2 | Hot, dry and smogy |
| ………. | ……………. | ………………. |
| 28/9/2009 | Cluster3 | dusty, fly ash, smogy, fog, Mist |
| 29/9/2009 | Cluster3 | dusty, fly ash, smogy, fog, Mist |
| 30/9/2009 | Cluster4 | hot, smogy and humid |
| ………. | ……….. | ……………….. |

Now, the accuracy of the above technique can be measured,

$$\text{Accuracy} = \frac{Number\ of\ matched\ records}{Total\ number\ of\ records} \times 100$$
$$= \frac{250}{300} \times 100$$
$$\cong 83.3\%$$

So, the accuracy of this method is 83.3%.

## VI. CONCLUSION AND FUTURE SCOPE

In this paper, a new technique is established to predict weather of upcoming days by the help of incremental K-means clustering algorithm. This technique is suitable for the dynamic databases where the climate data are changed frequently. In this paper the accuracy of this technique is calculated.

In future, other incremental clustering algorithms can be used to predict the weather and can compare them with each other to detect which algorithm among them provide better accuracy.

## ACKNOWLEDGMENT

Special thanks to Dr. S. Verma and Dr. T. S. Sinha from the National Institute of Technology and DIMAT Raipur, whose comments improved the presentation of this article.